\documentclass[11pt,a4paper]{article}
\usepackage{amssymb, amsmath, graphicx, subcaption, cite}
\usepackage{float}
\usepackage{graphpap,epsfig,textcomp}
\usepackage{epstopdf}
\usepackage{tabularx}
\usepackage[utf8]{inputenc}
\usepackage[english]{babel}
\usepackage{hyperref}
\usepackage[nottoc]{tocbibind}
\begin{document}

\begin{center}
{\Large {\bf Monte Carlo study of the phase transitions in the classical XY ferromagnets with random anisotropy}}\end{center}

\vskip 1cm

\begin{center}{\it Olivia Mallick$^1$ and Muktish Acharyya$^{2,*}$}\\
{\it Department of Physics, Presidency University,}\\
{\it 86/1 College Street, Calcutta-700073, INDIA}\\
{$^1$E-mail:olivia.rs@presiuniv.ac.in}\\
{$^2$E-mail:muktish.physics@presiuniv.ac.in}\end{center}

\vskip 1cm

\noindent {\bf Abstract:} The three-dimensional anisotropic classical XY ferromagnet has been investigated by extensive Monte Carlo simulation using the Metropolis single spin flip algorithm. The magnetization ($M$) and the susceptibility ($\chi$) are measured and studied as functions of the temperature of the system. For constant anisotropy, the ferro-para phase transition has been found to take place at a higher temperature than that observed in the isotropic case. The system gets ordered at higher temperatures for higher values of the strength of anisotropy. The opposite scenario is observed in the case of random anisotropy. For all three different kinds of statistical distributions (uniform, Gaussian, and bimodal) of random anisotropy, the system gets ordered at lower temperatures for higher values of the width of the distribution of anisotropy. We have provided the phase boundaries in the case of random anisotropy. The critical exponents for the scaling laws $M \sim L^{-{{\beta} \over {\nu}}}$ and $\chi \sim L^{{{\gamma} \over {\nu}}} $ are estimated through the finite size analysis.

\vskip 3cm


\vskip 2cm

\noindent $^*$ Corresponding author
\newpage

\noindent {\bf {I. Introduction:}}

The continuous symmetric classical spin models like XY are of great interest because they have very rich physical behaviours. The order parameter belonging to the XY model has only two components. For the first time, a new definition of order called topological order is proposed by Kosterlitz and Thouless for two-dimensional planar magnets like XY ferromagnets in which no long-range ferromagnetic order exists \cite{kosterlitz1,bishop,kosterlitz2}. Readers may consult the review article\cite{jose} to get a better understanding of the Berezinsky-Kosterlitz-Thouless (BKT) transition. Since then, the magnetic properties of the XY magnet have been studied with great interest. For example, the Monte Carlo simulation was carried out\cite{krech} to study the structure factor and transport properties of the three-dimensional XY model. The site -diluted classical XY model was investigated\cite{filho2} by extensive Monte Carlo simulation using a hybrid algorithm. It is found that the critical exponents and universal cumulants are independent of the amount of dilution. The Blume-Emery-Griffiths model has been investigated\cite{filho3} on thin films, considering the XY vectorial generalisation, and its thermodynamical properties are determined as a function of the film thickness. The surface critical behaviour of the XY model has been studied\cite{landau2} by Monte Carlo simulation to estimate the corresponding critical exponents. 

The effects, of the factor (anisotropy) responsible for breaking the SO(2) symmetry in the XY model, are summarized
here.
The anisotropic (with different interplane/intraplane coupling ratios) XY ferromagnet was investigated \cite{biplab} by the renormalization group technique to study the vortex-loop scaling behaviour. The renormalization group
technique has been employed\cite{granato} to study the critical behaviours of coupled XY model. The Cantor spectra have been observed\cite{satija} in the one dimensional quasiperiodic anisotropic XY model. 
The quantum version of the XY model in the presence of anisotropy is presented\cite{ma} in the boson space and
solved exactly in the spherical limit. 
The phase diagram is drawn in the plane formed by the temperature and anisotropy. 

A few studies on the dynamics of XY model are worth mentioning. The dynamics of the vortices were studied\cite{kim} by the time-dependent Ginzburg-Landau equation and estimated the dynamical critical exponent. The numerical integration of the dynamical equation of two dimensional XY model suggested\cite{yurke} the logarithmic growth of correlations. The Monte Carlo method was employed to investigate\cite{maccari} the interplay of spin waves and the vortices in two-dimensional XY ferromagnets at the limit of low vortex-core energy.

 Experimentally, the BKT transition was observed\cite{murthy} in ultracold Fermi gas. The numerical simulation was performed to study\cite{gouvea} the structure and dynamics of two-dimensional XY ferromagnets in the presence of an applied in-plane magnetic field. An extensive Monte Carlo simulation was employed\cite{univ,hasenbusch} to study the critical behaviour of a three-dimensional XY ferromagnet. The critical exponents are estimated and the universality class is found. Very recently, the phase transition was studied\cite{muktish} in the three-dimensional XY layered antiferromagnet.

The XY model of a quantum lattice fluid or a ferromagnet is studied\cite{betts}  by the method of
exact high-temperature series expansion. Analysis of these series yields the critical values $kT_c/J=4. 84\pm0. 06$,  $\gamma=1.00\pm0.07$, and
$\alpha=  - 0.20\pm0.20$ for the fcc lattice. The inverse energy cascade in the non-Kolmogorov energy spectrum has been observed\cite{tanogami} in a modified XY model.
The distribution of quantum coherence has been studied\cite{qin} in two dimensional XY model. The effects of transverse field in the XY model has been studied\cite{rosa}.
The quantum critical behaviour of spin-1/2 anisotropic XY model has been investigated
\cite{su} with staggered Dzyaloshinskii-Moriya interaction. 

XY model has recently been investigated for various kinds of interactions. Those are
higher order exchange\cite{zuko1},  antinematic
kind of interactions\cite{zuko3}, geometrically frustrated interaction\cite{lach1},
higher order antiferromagnetic interaction\cite{lach2} on a triangular lattice.The kinetics of glass transition in three dimensional XY model is analysed\cite{vasin} in the light of gauge
field theory.  Recently, the ageing and domain growth has been investigated\cite{puri} in random field
XY model.

How can the role of magnetic anisotropy be realized experimentally ?
The magnetic anisotropy can influence the Kondo effect and the spin transport. 
The magnetic anisotropy can split the single peak of differential conductance into two
peaks in the Kondo effect studied\cite{expt1} recently in single Co atom bound on top of a Cu atom
of the {${\rm Cu_2N}$} surface.
They observed that a Kondo resonance
emerges for large-spin atoms only when the magnetic anisotropy
creates degenerate ground-state levels that are connected by the
spin flip of a screening electron. The magnetic anisotropy also plays the major role
in determining how the Kondo resonance evolves in a magnetic field:
the resonance peak splits at rates that are strongly direction
dependent. This dependence on the directionality has been identified as the effect of magnetic anisotropy. Another example of the role of magnetic anisotropy in spin-filter junction may be referred
\cite{expt2} here. The magneto -
transport is largely determined by the magnetic anisotropy
at the interface or junction. The system is fabricated ${\rm LSMO/chromite/Fe_3O_4}$
 junctions where the chromite barrier layer, either ${\rm CoCr_2O_4}$
(CCO) or ${\rm MnCr_2O_4}$ (MCO), is isostructural with ${\rm Fe_3O_4}$ .
These ultrathin chromite layers exhibited normal
ferromagnetic behaviours below their bulk Curie temperature ($Tc$) and
proximity-induced ferromagnetism due to ${\rm Fe_3O_4}$ above their
bulk $T_c$ , thus giving rise to an effective spin-filter junction.
Although both chromite compounds form a normal spinel
structure with all $Cr^{3+}$ ions in the octahedral sites, the magnetic
anisotropy of the two compounds are opposite in sign consequently giving rise
to junction magnetoresistance values more than an
order of magnitude higher in CCO junctions compared to that of MCO
junctions.

In the brief introduction discussed above, although the effects of magnetic anisotropy are investigated both theoretically and experimentally, the systematic Monte Carlo study of the {\it critical behaviour} of anisotropic XY ferromagnets was not found in the literature. The isotropic XY ferromagnet has SO(2) symmetry (responsible for the 
appearance of vortex-antivortex pairing)  destroys the long range ferromagnetic ordering. The anisotropy breaks the SO(2) symmetry in XY ferromagnet. This broken SO(2) symmetry would lead to the destruction of vortex (and antivortex) with the  appearance of long range ferromagnetic ordering in the system. This should affect the ferro-para critical behaviour of the XY ferromagnet. What is the critical behaviour if anisotropy exists in the system? What does the phase boundary look like ? What are the critical exponents ? Above all, how does the critical behaviour will be affected if the anisotrpy is random ? Does this randomness affect the critical exponents ? What information can be obtained about the universality class of anisotropic XY ferromagnet ? These questions prompted us to study the anisotropic XY ferromagnet by Monte Carlo simulation. We have addressed these questions in our present study. We have considered three-dimensional anisotropic XY ferromagnet and studied the critical behaviour by extensive Monte Carlo simulation with Metropolis algorithm. We report the results of constant anisotropy and variable anisotropy (having specific statistical distributions), the phase boundaries and the corresponding critical exponents.  We believe, our results are new and significantly important in the field of phase transition in the continuous symmetric spin model ferromagnets with random disorder. These results will prompt the researchers to investigate the role of random anisotropy in the critical behaviours of continuous symmetric spin models.
The paper is organized as follows: In the next section (section II) the anisotropic XY ferromagnetic model is introduced, the Monte Carlo simulation scheme is discussed
briefly in section III. The numerical results are reported in section IV. The paper ends with concluding remarks in section V.
 
\vskip 3cm

\noindent {\bf {II. Anisotropic XY ferromagnetic model:}}

The  Hamiltonian of anisotropic XY ferromagnet is expressed as

\begin{equation}
H= -J \sum_{<i,j>} (1+\gamma_{ij})S_i^xS_j^x + (1-\gamma_{ij})S_i^yS_j^y
\end{equation}
where $S_{i}^x (=cos \theta_i)$ and $S_{i}^y(=sin \theta_i)$ correspond to  the components of the two dimensional spin vector (at the i-th lattice site) having unit length, $|S|$=1. $\theta$ is the angle (measured with respect to the positive X-axis)
of the spin vector. It may be considered a rotor at each lattice site, which can point in any direction in two dimensional XY plane.
$\gamma_{ij}$ is the anisotropy parameter. For $\gamma_{ij}=0$, the system becomes  isotropic XY ferromagnet and for $\gamma_{ij}=1$ the system represents ferromagnetic XX model (but $S_i^{x}$ is continuous variable).
$J$ is the nearest neighbour (represtented by $<ij>$ in the summation) ferromagnetic ($J > 0$) interaction strength. 
Anisotropy plays a crucial role in a magnetic model system. Thus we can study the effects of different types of anisotropy in the behaviours of magnetic model systems. Here we have considered mainly two different types of anisotropy.

\begin{itemize}
\item {\bf Constant Anisotropy:} Anisotropy is considered constant here for all spin-spin interactions. 

\item {\bf Distributed Anisotropy:} Anisotropy is considered as a random variable having three different types of statistical distribution.
\end{itemize}

\indent{a)\textit{Uniformly Distributed random anisotropy}}: The statistical distribution of variable anisotropy is uniform.

\begin{equation}
P_{u}(\gamma_{ij})={1\over \omega}, 
\label{uniform}
\end{equation}

\noindent where $\omega$ is the width of the distribution. All values of anisotropy in the range $-\omega/2$ to $+\omega/2$ are randomly quenched on each lattice site.The standard deviation of the distribution is $\sigma_{u}={\omega\over{2\sqrt{3}}}$.
 
\indent{b)\textit{ Normally Distributed random anisotropy}}: Variable anisotropy having the Gaussian probability distribution,

\begin{equation}
P_{n}(\gamma_{ij})={1\over{\sqrt{2\pi}\omega}}e^{-{\gamma_{ij}^2\over{2{\omega^2}}}}.
\label{normal}
\end{equation}
\noindent This distribution is obtained by using Box-Muller algorithm\cite{box}. 
Here $\omega$, the width of the distribution, is equal to standard deviation $\sigma_n$.

\indent{c)\textit{ Bimodal Distribution of random anisotropy}}: Bimodal distribution of variable anisotropy,
\begin{equation}
P_{b}(\gamma_{ij})={1\over 2}[\delta(\gamma_{ij}-{\omega\over 2})+\delta(\gamma_{ij}+{\omega\over 2})].
\label{bimodal}
\end{equation}
\noindent where $\delta$ represents the Dirac Delta function. This distribution implies that, 
the values of the anisotropy can be either $-\omega/2$ or $+\omega/2$ with equal probability.  Here, the standard deviation relates to width of the distribution as $\sigma_b={\omega\over 2}$.

It is worth mentioning that for each of the above statistical distributions, the mean value of anisotropy $<\gamma_{ij}>$=0.

\vskip 3cm

\noindent {\bf {III. Monte Carlo simulation method:}}

A three dimensional simple cubic lattice of size $L$
(=20 here, apart from the finite size analysis)
is considered here. It may be noted here that the spatial dimensions of the 
system (lattice) are three
(cubic) and the dimensions of the spin vector are two (XY model). The boundary 
conditions are taken as periodic in all three directions of the lattice.

The simulation starts from a random initial spin configuration corresponding to
a very high temperature phase. This corresponds to the paramagnetic phase having
zero magnetisation.
At any finite temperature $T$ (measured in the unit of $J/k$, where $k$ is Boltzmann
constant), a site (say x,y,z) is chosen randomly 
(at any instant $t$) having an initial spin configuration (represented
by an angle $\theta_i(x,y,z,t)$). A new configuration of the spin 
(at site x,y,z and at the same instant $t$) is also chosen
(represented by $\theta_f(x,y,z,t)$) randomly. The change in energy
($\delta H(t)$) due to the change in configuration (angle)
of spin (from $\theta_i(x,y,z,t)$ to $\theta_f(x,y,z,t)$) is calculated 
from equation (1). The probability of accepting
the new configuration (in the next time) is calculated from the Metropolis formula\cite{binder,simulation}

\begin{equation}
P_f = {\rm Min}[{\rm exp}({{-\delta H(t)} \over {kT}}), 1].
\end{equation}

A uniformly  distributed (between 0 and 1) random number ($r=[0,1]$) is chosen. 
The chosen site is assigned to the new
spin configuration $\theta_f(x,y,z,t')$ (for the next instant $t'$)
if $r \leq P_f$. In this way, $L^3$ number of sites are updated randomly. $L^3$ number of such
random updates defines a unit time step and is called Monte Carlo step per site (MCSS). The time in this simulation is measured in the unit of MCSS.
Throughout the study the system size $L(=20)$.  The total length of simulation is $2\times10^4$ MCSS, out of which the initial
$10^4$ MCSS times are discarded. Here, to make the system ergodic some initial MCSS are required. All statistical quantities are calculated by averaging over the rest $10^4$ MCSS\cite{binder}. It may be noted here that
 the measurements for the magnetisation at each MCSS are
correlated. These correlations should not be ignored as they can have a serious effect on the critical behaviour studied
in this paper.

\vskip 0.3cm

The instantaneous components of magnetisations are 
$M_{x}(t)= {1 \over {L^3}} \sum s_x (x,y,z,t)\\ 
= {1 \over {L^3}} \sum {\rm cos} (\theta (x,y,z,t))$ 
and 
$M_{y}(t)= {1 \over {L^3}} \sum s_y(x,y,z,t)
= {1 \over {L^3}} \sum {\rm sin}(\theta(x,y,z,t)$.

The equilibrium magnetisation is measured as 
$M=\sqrt{M_{x}^2+M_{y}^2}$.
The susceptibility is determined by 
$\chi={L^3\over{kT}}(<M^2>-<M>^2)$. The symbol $<..>$, represents the time averaging,
which is approximately (within the length of simulation) equal to the ensemble averaging in the ergodic limit.

It may be noted here, the values of MCSS required for averaging are quite large for larger system sizes (say $L$=25, 30 etc.
we have used for finite size analysis). For $L=30$, the initial 27000 MCSS are discarded and calculated the averaged 
quantities over further 8000 MCSS. Moreover, for random anisotropy the time average quantities are further averaged over
many (ranging from 10 to 200 depending on the values of temperature and system size) disorder realizations. The more samples (different random disorder realizations) are required near the transition temperature.

\vskip 1.0 cm

\noindent {\bf {IV. Simulational Results:}}

\vskip 0.5cm
In the following, we will report the results of Monte Carlo study for the three-dimensional anisotropic XY model for different types of the distribution of anisotropy. 

\noindent {\bf {\it (a). Constant Anisotropy}}

We will first consider that anisotropy ($\gamma_{ij}=\gamma ~~\forall ~~i,j$) is constant.  We have started with a high temperature random configuration  of the spins. This corresponds to the
paramagnetic phase. The system is slowly cooled down to achieve the equilibrium 
state of the system and then we have calculated the various thermodynamic quantities. Here the phrase “slowly cooled down” means the system is being cooled with small step
sizes (starting from a high temperature random paramagnetic configuration). Here, we have
taken the stepsize $dT$=0.05. It may be noted here, we have reduced ($dT=0.02$) the stepsize near the transition temperature. We have not used simulated annealing.
In such
cooling process, the final spin configuration for any temperature has been considered
the initial spin configuration of immediate lower temperature. We have studied the variations of magnetisation ($M$) and susceptibility ($\chi$) as the 
functions of the temperature of the system for different strengths of anisotropy.

\begin{figure}[h]
\begin{center}

%\resizebox{10cm}{!}{\includegraphics[angle=0]{M-T-different-const-anis.pdf}}

\resizebox{17cm}{!}{\includegraphics[angle=0]{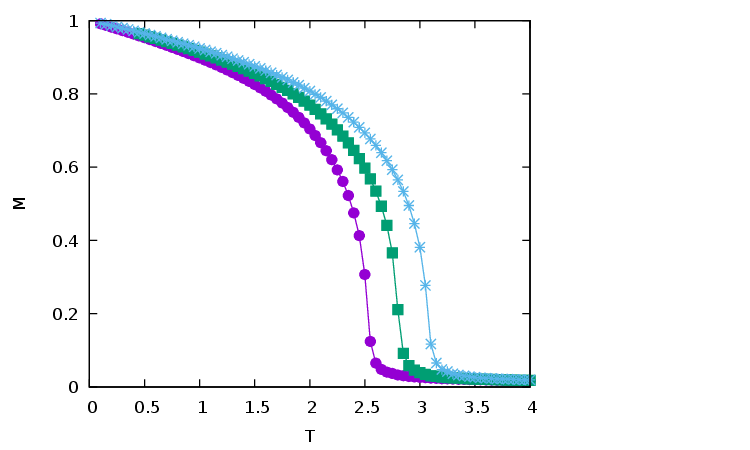}}

\caption{~Caption:~The magnetisation ($M$) is plotted against the Temperature ($T$) for different values of the strength of the constant anisotropy ($\gamma$). Here $\gamma=0.1$(Bullet), $\gamma=0.2$(Box), $\gamma=0.3$(Star).\\
Figure-1:~Alt-text: ~The temperature dependence of the magnetisation for three different strengths of the constant anisotropy. The phase transition (where the magnetisation vanishes) takes place at higher temperature for higher value of anisotropy.
}

\label{M-T-different-const-anis}
\end{center}
\end{figure}

\begin{figure}[h]
\begin{center}

%\resizebox{10cm}{!}{\includegraphics[angle=0]{Chi-T-different-const-anis.pdf}}

\resizebox{17cm}{!}{\includegraphics[angle=0]{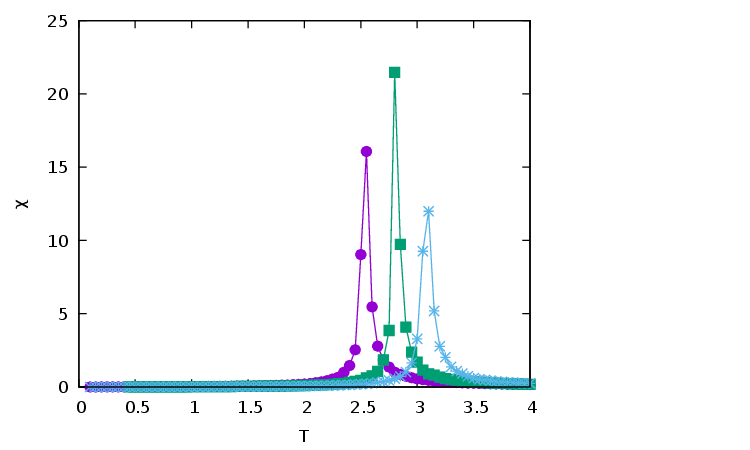}}

\caption{~Caption:~The susceptibility ($\chi$) is plotted against the Temperature ($T$) for different values of the strength of the constant anisotropy ($\gamma$). Here $\gamma=0.1$(Bullet), $\gamma=0.2$(Box), $\gamma=0.3$(Star).\\
Figure-2:~Alt-text:~The temperature dependence of the susceptibility for three different strengths of the constant anisotropy. The susceptibility gets peaked at pseudocritical temperature indicating the phase transition. The phase transition takes place at higher temperature for higher value of the anisotropy.}
\label{Chi-T-different-const-anis}
\end{center}
\end{figure}

\begin{figure}[h]
\begin{center}

%\resizebox{10cm}{!}{\includegraphics[angle=0]{Tc-gamma-const-anis.pdf}}

\resizebox{17cm}{!}{\includegraphics[angle=0]{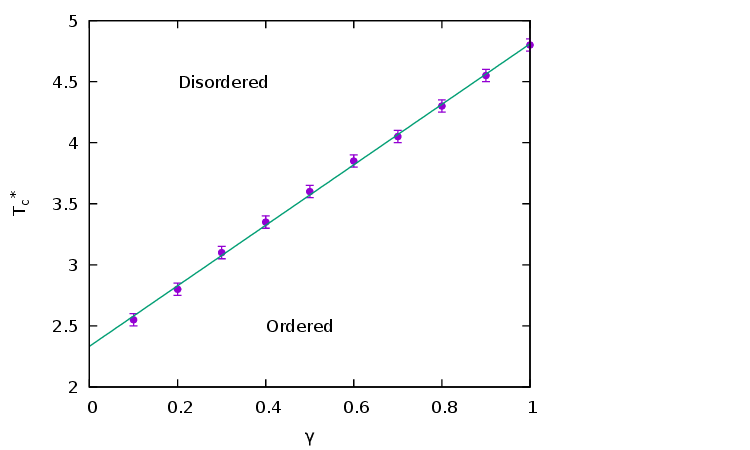}}

\caption{~Caption:~The pseudocritical temperature ($T_c^{*}$)is plotted against the strength 
of the constant anisotropy ($\gamma$). The fitted straight line is $T_c^{*}=
a\gamma +b$, where, $a=2.4757\pm0.0289$, $b=2.3333\pm0.0179$ and $\chi^2=0.0262$.\\
Figure-3:~Alt-text:~ The variation of pseudocritical temperature with the strength of the constant anisotropy, is shown here. The pseudocritical temperature increases as the anisotropy increases. The pseudocritical temperature varies linearly with the strength of the constant anisotropy.}
\label{Tc-gamma-const-anis}
\end{center}
\end{figure}

As the temperature decreases the magnetisation grows and is  
found to saturate eventually to unity . 
In Fig-\ref{M-T-different-const-anis} and Fig-\ref{Chi-T-different-const-anis}, the MC numerical results for the thermodynamic quantities $M$ and $\chi$ are reported
respectively for different values of constant anisotropy  $\gamma$=0.1,0.2,0.3. The peak of the susceptibility 
 reflects the existence of conventional ferro to para phase transition for three dimensional  anisotropic XY Model. From the position of 
 the peak of the susceptibility, we obtain the pseudocritical temperature (this is the critical temperature for finite  system) for the ferro-para phase transition. This pseudocritical temperature will eventually achieve the true critical temperature in the thermodynamic limit i.e., $L \to \infty$. The pseudocritical temperature is found to
 increase with the  strength of uniform anisotropy.
 The pseudocritical temperature increases linearly with the strength of anisotropy increases. In Fig-\ref{Tc-gamma-const-anis}, from the straight line fit we estimate the value for pseudocritical temperature $T_{c} =2.33J/k$ for $\gamma=0$. This result can be compared
 with the previous MC estimate\cite{univ,hasenbusch} of $T_c$ for {\it isotropic} XY
 ferromagnet. Moreover, for $\gamma=1$, the system maps to XX model. The pseudocritical
 temperature for $\gamma=1$ is closer to that for the MC estimate ($T_c=4.511..$)\cite{binder,simulation} of 3D Ising ferromagnet.

\begin{figure}[h]
\begin{center}

%\resizebox{10cm}{!}{\includegraphics[angle=0]{Binder-T-diff-L-const-anis.pdf}}

\resizebox{17cm}{!}{\includegraphics[angle=0]{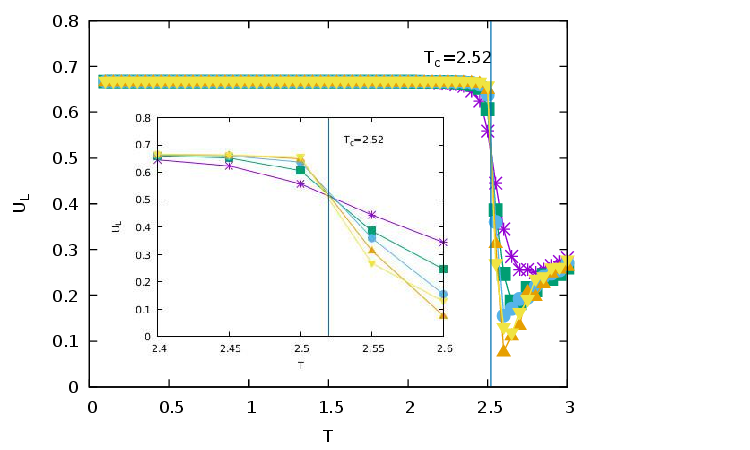}}

\caption{~Caption:~The fourth order Binder cumulant ($U_L$) is plotted against the Temperature ($T$) for different values of $L$. Here $L=10$(Star), $L=15$(Box), $L=20$(Bullet), $L=25$(Triangle) and $L=30$(Inverted Triangle). The vertical line passing through the common intersection represents the critical temperature $T_c=2.52J/k$. The strength  of the constant anisotropy is $\gamma=0.1$. The inset shows the enlarged view to have a clear vision
near the common intersection.\\
Figure-4:~Alt-text:~ The temperature dependence of the fourth order 
Binder cumulant for five different system sizes. The common intersection of all curves indicates the critical temperature. The vertical line, passing through the common intersection, 
estimates the critical temperature. Here the anisotropy is constant. The enlarged
view (near the common intersection) is shown in the inset.}
\label{Binder-T-diff-L-const-anis}
\end{center}
\end{figure}

\begin{figure}[h]
\begin{center}

%\resizebox{10cm}{!}{\includegraphics[angle=0]{M-T-diff-L-const-anis.pdf}}

\resizebox{17cm}{!}{\includegraphics[angle=0]{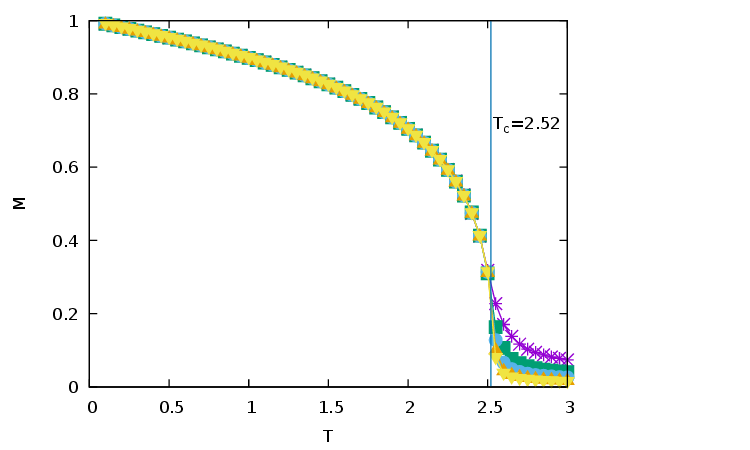}}

\caption{~Caption:~The magnetisation ($M$) is plotted against the temperature ($T$) for different system size ($L$) and a fixed strength 
of the constant anisotropy ($\gamma$=0.1). Here $L=10$(Star), $L=15$(Box), $L=20$(Bullet), $L=25$(Triangle) and $L=30$(Inverted Triangle). The vertical line represents the critical temperature $T_c=2.52J/k$. \\
Figure-5:~Alt-text:~The temperature dependence of the magnetisation for five different sizes of the system for a fixed value of constant anisotropy. The vertical line indicates the critical temperature. The value of the magnetisation (for any system size) at the critical
temperature is found from the intersection with the vertical line. The critical magnetisation decreases as the system size increases. This is the indication of the finite size effect of thermodynamic phase transition.}
\label{M-T-diff-L-const-anis}
\end{center}
\end{figure}

\begin{figure}[h]
\begin{center}

%\resizebox{10cm}{!}{\includegraphics{M-Tc-L-const-anis.pdf}}

\resizebox{17cm}{!}{\includegraphics[angle=0]{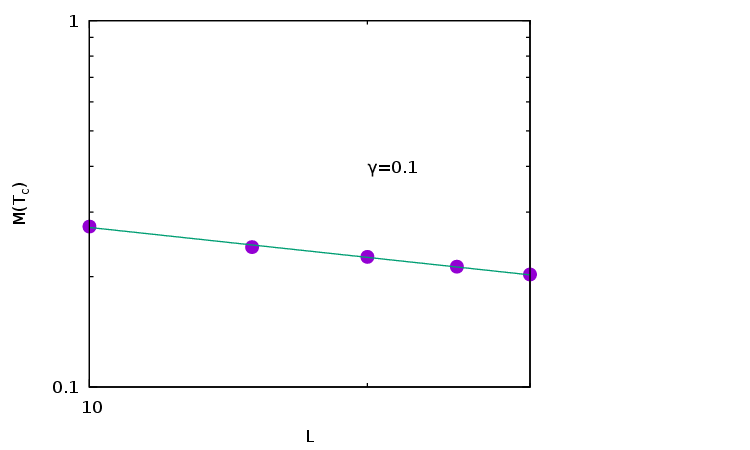}}

\caption{~Caption:~The magnetisation at $T_c$ ($M(T_c)$) are plotted against the system size ($L$) in logarithmic scale. The fitted line is
$M(T_c)=aL^{-{{\beta} \over {\nu}}}$. Here, $a=0.509\pm0.015$ and ${{\beta} \over {\nu}}=0.271\pm0.010$. The value of $\chi^2=0.0021$. The strength of the constant anisotropy  is 0.1.\\
Figure-6:~Alt-text:~The size dependence of the magnetisation at critical temperature for constant anisotropy shown in the logarithmic scale for five different system sizes. As the
size increases the magnetisation (at the critical temperature) decreases in a power law fashion (linear graph in logarithmic scale). This indicates that the critical magnetisation vanishes in the thermodynamic limit.}
\label{M(Tc)-L-const-anis}
\end{center}
\end{figure}

The critical temperature of the ferro-para phase transition of anisotropic (constant) XY ferromagnet, in the thermodynamic limit (L{$\rightarrow$} $\infty$) is determined by considering the thermal variation of fourth order Binder Cumulant $U_{L}(T)=1-{<m^4>_{L}\over{3<m^2>^2_{L}}}$ of magnetisation for different system sizes. The variation of Binder cumulant $U_{L}(T)$ with temperature for different system sizes is depicted in the Fig-\ref{Binder-T-diff-L-const-anis}. In this case, the strength of anisotropy is taken $\gamma=0.1$. From the common intersection of Binder cumulant of different system sizes, we obtained the value of critical temperature as $T_{c}=2.52J/k$. The variation of magnetisation ($M$)  with temperature for different system sizes ($L=10,15,20,25,30$) has been shown in Fig-\ref{M-T-diff-L-const-anis}. At the critical
temperature $T_c$, the magnetisation $M$ should vanish. But for finite sized system
$M$ assumes a nonzero value. However, in the thermodynamic limit ($L \to \infty$),
$M$ must vanish. We have also studied this so called finite size analysis. Assuming,
the scaling relation, $M(T_c) \sim L^{-{\beta\over\nu}}$, the exponent,
${\beta\over\nu}$ has been estimated (Fig-\ref{M(Tc)-L-const-anis}) as equal to $0.271\pm0.010$. 

\begin{figure}[h]
\begin{center}

%\resizebox{10cm}{!}{\includegraphics{Chi-T-diff-L-const-anis.pdf}}

\resizebox{17cm}{!}{\includegraphics[angle=0]{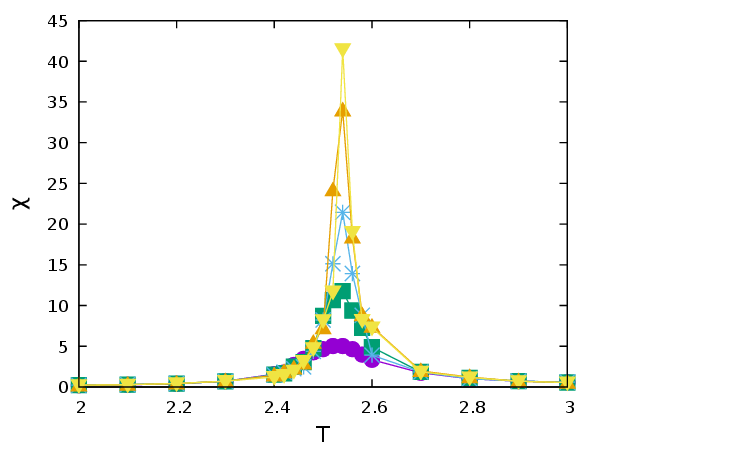}}

\caption{~Caption:~The susceptibility ($\chi$) is plotted against the temperature ($T$) for different system size ($L$) and a fixed strength 
of the constant anisotropy ($\gamma$=0.1).Here $L=10$(Bullet), $L=15$(Box), $L=20$(Star), $L=25$(Triangle) and $L=30$(Inverted Triangle).\\
Figure-7:~Alt-text:~Temperature dependence of the susceptibility for five different system sizes and for a fixed value of the constant anisotropy. The peak height of the susceptibility increases as the system size increases. This is indicating the finite size effect of the thermodynamic phase transition.}
\label{Chi-T-diff-L-const-anis}
\end{center}
\end{figure}

\begin{figure}[h]
\begin{center}

%\resizebox{10cm}{!}{\includegraphics{Chi-peak-L-const-anis.pdf}}

\resizebox{17cm}{!}{\includegraphics[angle=0]{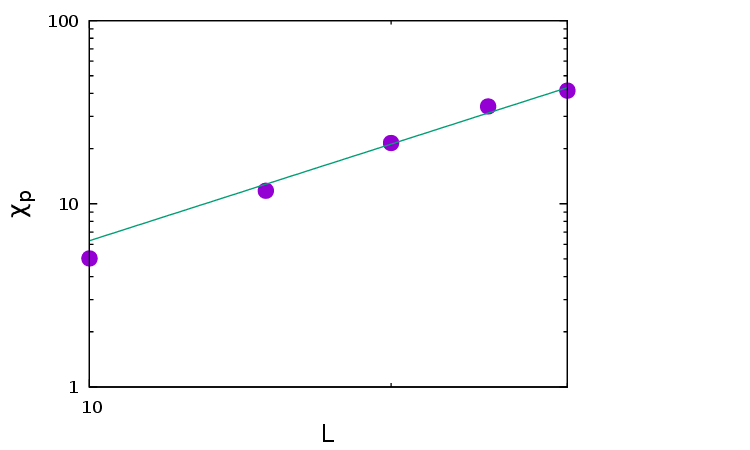}}

\caption{~Caption:~The peak values of the susceptibilities ($\chi_p$) are plotted against the system size ($L$) in logarithmic scale. The fitted line is
$\chi_p=aL^{{\gamma} \over {\nu}}$. Here, $a=0.111\pm0.059$, 
${{\gamma} \over {\nu}}=1.751\pm0.162$ and $\chi^2=4.15$. The strength  of the constant anisotropy $\gamma=0.1$.\\
Figure-8:~Alt-text:~The system size dependence of the maximum value of the susceptibility for constant anisotropy is shown here in logarithmic scale for five different system sizes and for a fixed value of constant anisotropy.
The maximum value of the susceptibility increases as the system size increases. The linear graph (in logarithmic scale) indicates that the susceptibility will eventually diverge in a power law fashion in the thermodynamic limit.
}
\label{Chi-peak-L-const-anis}
\end{center}
\end{figure}

In continuation of the finite size analysis, the susceptibility ($\chi$) has been
studied as a function of the temperature ($T$) for different system sizes ($L$).
The peak height of the susceptibility ($\chi$) is observed to 
increase (Fig-\ref{Chi-T-diff-L-const-anis})
as the system
size ($L$) increased. This is indeed the true picture of the thermodynamic phase
transition in the thermodynamic limit ($L$). A remarkable feature of equilibrium
phase transition indicates an important fact of growth of critical 
correlations\cite{stanley}. Since, the susceptibility is directly related to the correlation function, the 
growth of the susceptibility near the critical point is a manifestation of the growth of correlation nera the critical point.
In the thermodynamic limit the susceptibility ($\chi$)
diverges at the critical temperature ($T_c$). Assuming the scaling relation,
for the peak value ($\chi_p$) of the susceptibility ($\chi$), $\chi_p\sim L^{\gamma\over\nu}$, the exponent ${\gamma\over\nu}$ has been estimated as
equal to $1.751\pm0.162$. This has been shown in Fig-\ref{Chi-peak-L-const-anis}.

What will happen if the strength of the anisotropy is random ?
 This has been studied for three different kinds of statistical distributions
of the random anisotropy  and the results are discussed in the next
subsections.

\vskip 0.5cm

\noindent {\bf {\it {(b). Distributed Anisotropy}}}

\vskip 0.3cm

\noindent {\it {(i) Uniformly distributed random anisotropy}}

\begin{figure}[h]
\begin{center}

%\resizebox{10cm}{!}{\includegraphics{M-T-different-uniform-anis.pdf}}

\resizebox{17cm}{!}{\includegraphics[angle=0]{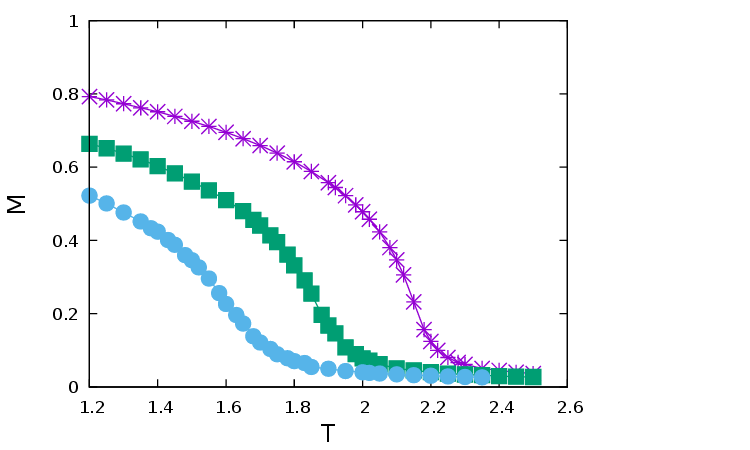}}

\caption{~Caption:~The magnetisation ($M$) is plotted against the temperature ($T$) for different widths of  uniform distribution of random anisotropy ($\gamma_{ij}$). Here $\omega=1.0$(Star), $\omega=2.5$(Box), $\omega=3.5$(Bullet).\\
Figure-9:~Alt-text:~The temperature dependence of magnetisation for three different
widths of the uniform distribution of random anisotropy. The phase transition (where the magnetisation vanishes) takes place at lower temperature for higher
value of the width of the uniform distribution of random anisotropy.}
\label{M-T-different-uniform-anis}
\end{center}
\end{figure}

\begin{figure}[h]
\begin{center}

%\resizebox{10cm}{!}{\includegraphics{Chi-T-different-uniform-anis.pdf}}

\resizebox{17cm}{!}{\includegraphics[angle=0]{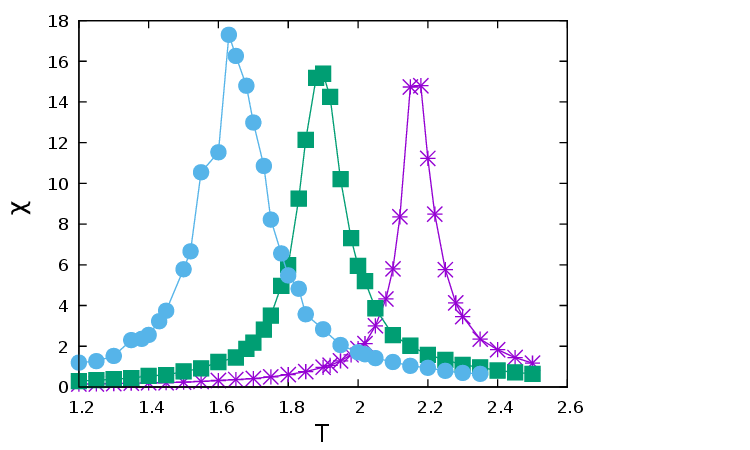}}

\caption{~Caption:~The susceptibility ($\chi$) is plotted against the Temperature ($T$) for different widths of  uniform  distribution of the random anisotropy ($\gamma_{ij}$). Here $\omega=1.0$(Star), $\omega=2.5$(Box), $\omega=3.5$(Bullet).\\
Figure-10:~Alt-text:~The temperature dependence of the susceptibility for three
different widths of the uniform distribution of random anisotropy. The susceptibility gets peaked (where the phase transition
takes place) at lower temperature
for higher value of the width of the uniform distribution of random anisotropy.}
\label{Chi-T-different-uniform-anis}
\end{center}
\end{figure}

Let us consider a random uniform distribution of anisotropy. The values of the anisotropy($\gamma _{ij})$ are uniformly distributed between $-\omega/2$ to $+\omega/2$. The probability distribution function for such
distribution is given in the equation-\ref{uniform}. The growth of 
magnetic order by cooling the system from a high temperature 
disordered phase is studied here. We have shown the thermal variation of the magnetisation (for different values of the width $\omega$ of random uniform distribution of the strength of anisotropy) in Fig-\ref{M-T-different-uniform-anis}.
The results show that the system starts to be ordered at lower temperatures for
higher values of the width of the distribution of anisotropy. Another feature may
also be noticed that the value of the zero temperature saturation magnetisation 
gets reduced more for the higher values of the width of the distribution. The
reduction of the pseudocritical temperature by the distributed anisotropy
 is clearly observed in the study of
the susceptibility ($\chi$) as a function of the temperature ($T$) of the system.
Fig-\ref{Chi-T-different-uniform-anis} shows the  variation of  susceptibility with temperature for three different widths ($\omega$=1.0,2.5,3.5) of the distribution
of the anisotropy. We have estimated the pseudocritical temperature of phase transition from the position of the peak of the susceptibility, for any value of the width of the distribution.  We observed that the peak shifts towards the left (lower temperature), indicating that critical temperature decreases as the width of the distribution increases. 

\begin{figure}[h]
\begin{center}

%\resizebox{10cm}{!}{\includegraphics{Binder-T-diff-L-uniform-anis.pdf}}

\resizebox{17cm}{!}{\includegraphics[angle=0]{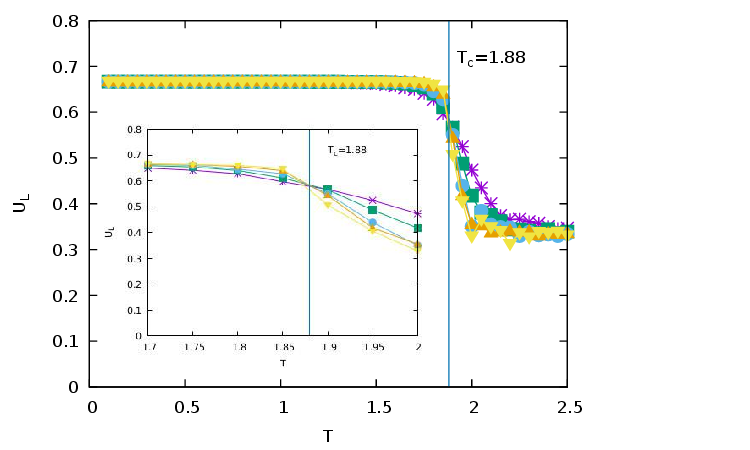}}

\caption{~Caption:~The fourth order Binder cumulant ($U_L$) is plotted against the Temperature ($T$) for different values of $L$. Here $L=10$(Star), $L=15$(Box), $L=20$(Bullet), $L=25$(Triangle) and $L=30$(Inverted Triangle). The vertical line passing through the common intersection represents the critical temperature $T_c=1.88J/k$. The width of the uniform distribution of the 
random anisotropy ($\gamma_{ij}$) is $\omega=2.5$.The inset shows the enlarged view to have a clear vision
near the common intersection.\\
Figure-11:~Alt-text:~Temperature dependence of the fourth order Binder cumulant for five different system sizes and for fixed width of the uniform distribution of random anisotropy.The common intersection of all curves is the critical temperature. The vertical line, passing through the common intersection,
estimates the critical temperature.  The enlarged
view (near the common intersection) is shown in the inset. }
\label{Binder-T-diff-L-uniform-anis}
\end{center}
\end{figure}

In the simulation, we have considered the lattices with a finite size ($L$) of
the system. However, the results in the thermodynamics limit 
($L \rightarrow$ $\infty$) have to be extracted. 
For this purpose, the finite size analysis  has been carried out by varying the system size $L=10,15,20,25,30$ for the width of the distribution $\omega=2.5$. 
To determine the actual transition temperature in the thermodynamic limit the fourth
order Binder cumulant ($U_{4}(L)$) is studied as function of the temperature for different 
sizes ($L$)
of the system.
The variation of the fourth order Binder cumulant, with temperature for different system sizes ($L$) is shown in Fig-\ref{Binder-T-diff-L-uniform-anis}. We 
have estimated the critical temperature $T_{c}=1.88J/k$ of the phase transition from the common intersection of Binder cumulant.

\begin{figure}[h]
\begin{center}

%\resizebox{10cm}{!}{\includegraphics{M-T-diff-L-uniform-anis.pdf}}

\resizebox{17cm}{!}{\includegraphics[angle=0]{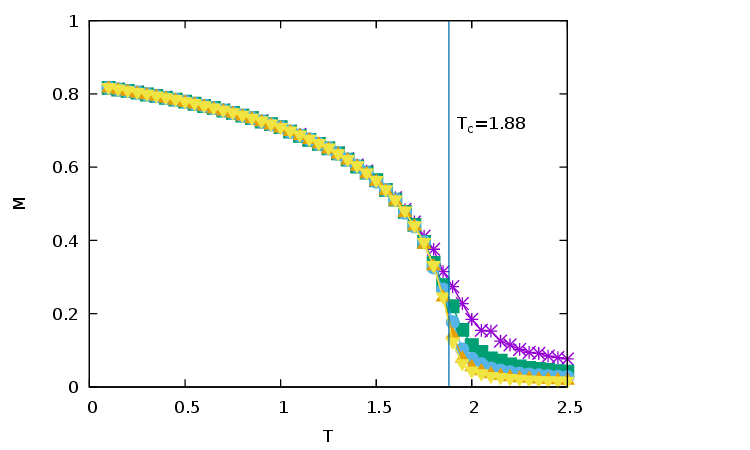}}

\caption{~Caption:~The magnetisation ($M$) is plotted against the temperature ($T$) for different system size ($L$) and a fixed width ($\omega=2.5$)of 
the  uniform  distribution of the 
random anisotropy ($\gamma_{ij}$). Here $L=10$(Star), $L=15$(Box), $L=20$(Bullet), $L=25$(Triangle) and $L=30$(Inverted Triangle). The vertical line represents the critical temperature $T_c=1.88J/k$.\\
Figure-12:~Alt-text:~Temperature dependence of magnetisation for five different system sizes and for a fixed width of the uniform distribution of random anisotropy. The vertical line indicates the critical temperature. The value of the magnetisation (for any system size) at the critical
temperature is found from the intersection with the vertical line. This critical magnetisation  decreases as the system size increases indicating the finite size effect of thermodynamic phase transition.
}
\label{M-T-diff-L-uniform-anis}
\end{center}
\end{figure}

\begin{figure}[h]
\begin{center}

%\resizebox{10cm}{!}{\includegraphics{M-Tc-L-uniform-anis.pdf}}

\resizebox{17cm}{!}{\includegraphics[angle=0]{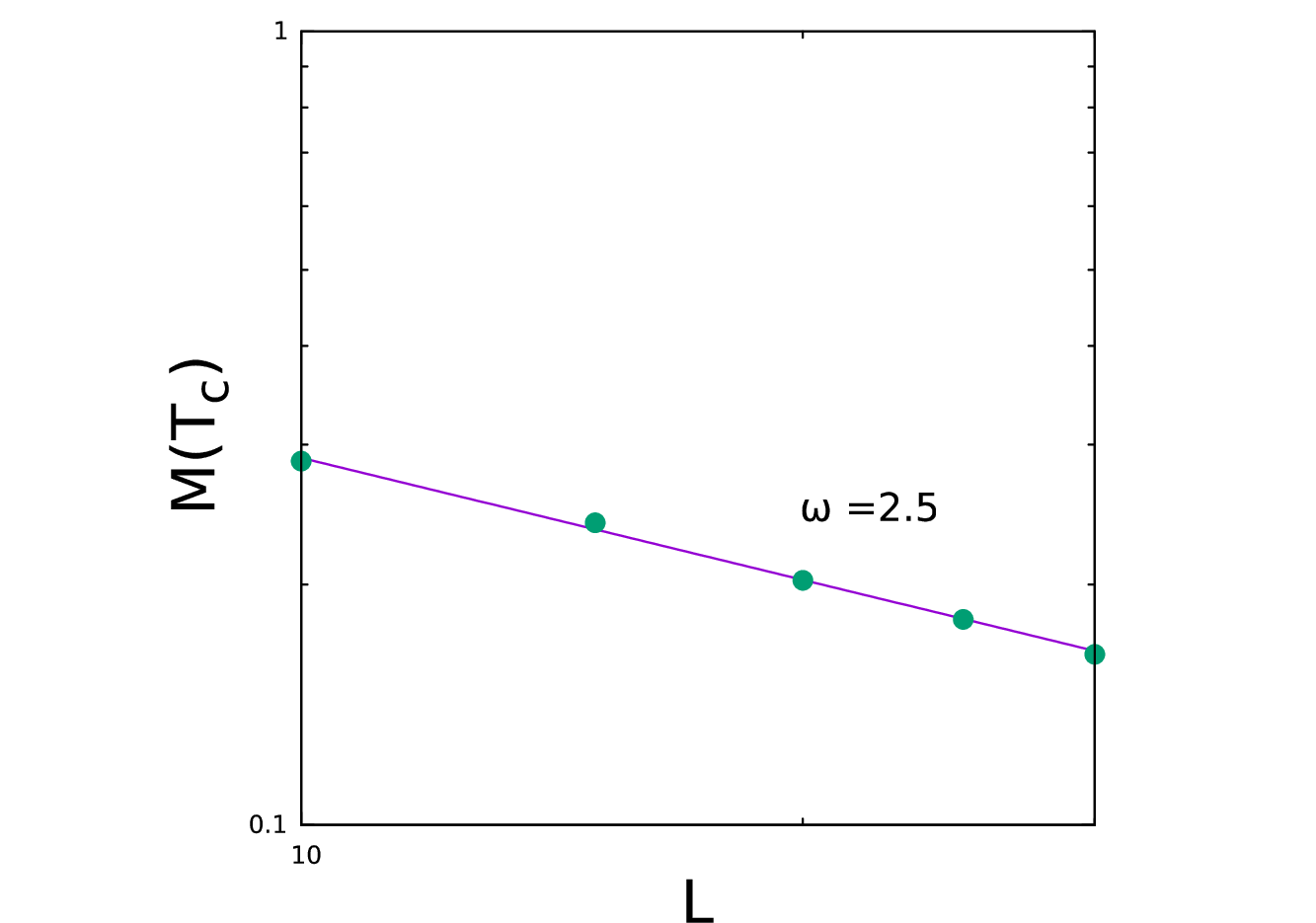}}

\caption{~Caption:~The magnetisation at $T_c$ ($M(T_c)$) are plotted against the system size ($L$) in logarithmic scale. The fitted line is
$M(T_c)=aL^{-{{\beta} \over {\nu}}}$. Here, $a=0.935\pm0.040$ and ${{\beta} \over {\nu}}=0.510\pm0.015$. The value of $\chi^2=0.00002$. The width of the uniform distribution  of the random anisotropy ($\gamma_{ij}$) is $\omega$=2.5 here. \\
Figure-13:~Alt-text:~The system size dependence (shown in logarithmic scale for five different system sizes) of the magnetisation at the
critical temperature for a fixed width of the uniform distribution of random anisotropy. This indicates that the critical magnetisation vanishes in the thermodynamic limit. The critical magnetisation decreases in a power law fashion (linear graph
in logarithmic scale) as the system size increases. The critical magnetisation vanishes in the thermodynamic limit.
}
\label{M(Tc)-L-uniform-anis}
\end{center}
\end{figure}

What will be the value of the magnetisation at $T_c=1.88J/k$ for different values
of the sizes of the system ? We have studied the magnetisation ($M$) as function of the temperature for different sizes ($L$) of the system and for a fixed value
($\omega=2.5$) of the width of the distribution of anisotropy. The results are shown in 
Fig-\ref{M-T-diff-L-uniform-anis}. In the thermodynamic limit, the magnetisation should 
vanish at $T_c$. The values of the magnetisation at $T_c=1.88J/k$ for different
$L$ are obtained from the method of linear interpolation. These values of the
magnetisation are studied as function of $L$. The finite size study, assuming the scaling relation, $M(T_c) \sim L^{-{\beta\over\nu}}$, yields ${\beta\over\nu}=0.510\pm0.015$ (Fig-\ref{M(Tc)-L-uniform-anis}).

\begin{figure}[h]
\begin{center}

%\resizebox{10cm}{!}{\includegraphics{Chi-T-diff-L-uniform-anis.pdf}}

\resizebox{17cm}{!}{\includegraphics[angle=0]{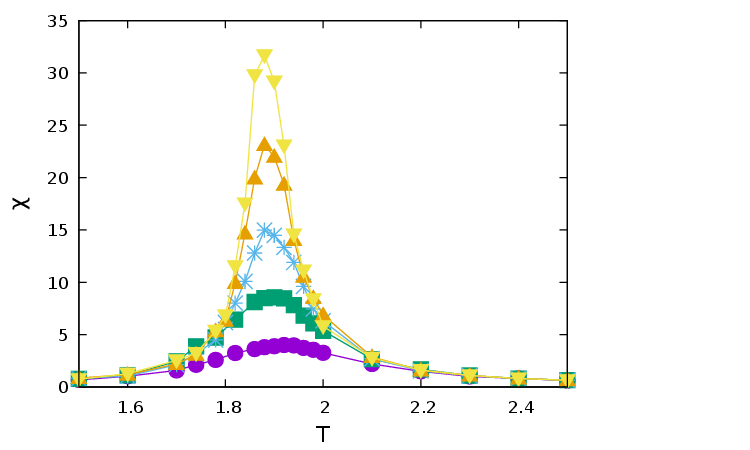}}

\caption{~Caption:~The susceptibility ($\chi$) is plotted against the temperature ($T$) for different system size ($L$) and a fixed width ($\omega$=2.5) of the uniform  distribution of the random anisotropy ($\gamma_{ij}$).Here $L=10$(Bullet), $L=15$(Box), $L=20$(Star), $L=25$(Triangle) and $L=30$(Inverted Triangle).\\
Figure-14:~Alt-text:~The temperature dependence of the susceptibility for
five different system sizes and for a fixed width of the uniform distribution of the random
anisotropy. The peak height of the susceptibility increases as the system size increases. This indicates the finite size effect
of the thermodynamic phase transition.}
\label{Chi-T-diff-L-uniform-anis}
\end{center}
\end{figure}

\begin{figure}[h]
\begin{center}

%\resizebox{10cm}{!}{\includegraphics{Chi-peak-L-uniform-anis.pdf}}

\resizebox{17cm}{!}{\includegraphics[angle=0]{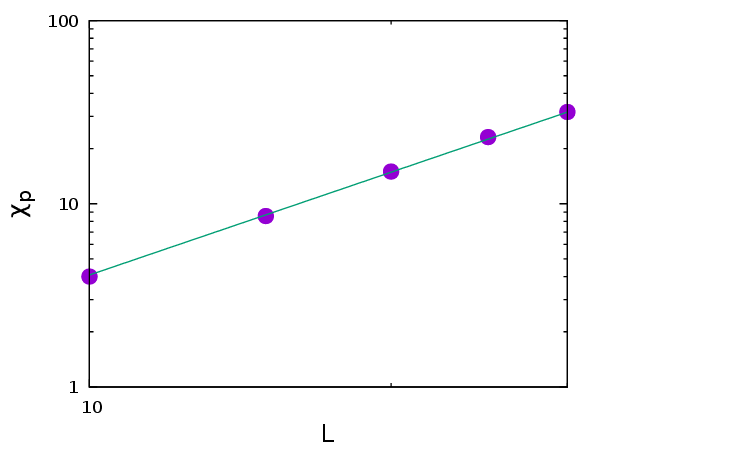}}

\caption{~Caption:~The peak values of the susceptibilities ($\chi_p$) are plotted against the system size ($L$) in logarithmic scale. The fitted line is
 $\chi_p=aL^{{\gamma} \over {\nu}}$. Here, $a=0.056\pm 0.006$, 
 ${{\gamma} \over {\nu}}=1.863\pm 0.033$ and $\chi^2=0.2949$.  The width of the uniform distribution  of the random anisotropy is $\omega$=2.5 .\\
 Figure-15:~Alt-text:~The system size dependence of the maximum value of the susceptibility, for five different system sizes and for a fixed width of the uniform distribution of the random anisotropy, is shown in logarithmic scale. The maximum value of the susceptibility increases  as the system size increases.  The linear graph (in logarithmic scale) indicates that the critical susceptibility will eventually diverge in a power law fashion in the thermodynamic limit.}
\label{Chi-peak-L-uniform-anis}
\end{center}
\end{figure}

This phase transition is supported by the growth of critical correlation. The
susceptibility ($\chi$) has been studied as function of the temperature ($T$) for
different sizes ($L$) of the system and a fixed value of the width ($\omega$) of
the distribution of the anisotropy. Fig-\ref{Chi-T-diff-L-uniform-anis} shows the
results. From the figure, it is clearly observed that the height of the peak of
the susceptibility ($\chi$) increases as $L$ increases. How does it diverge at
the critical temperature ? This has been studied by plotting the values of the 
height of the peak ($\chi_p$)
of the susceptibility as function of $L$. 
The plot (Fig-\ref{Chi-peak-L-uniform-anis}) in the 
logarithmic scale supports the scaling relation $\chi_p\sim L^{\gamma\over\nu}$ with
estimated value ${\gamma\over\nu}=1.836\pm0.033$.

\noindent {\it {(ii) Normally distributed random anisotropy}}

\begin{figure}[h]
\begin{center}

%\resizebox{10cm}{!}{\includegraphics{M-T-different-gaussian-anis.pdf}}

\resizebox{17cm}{!}{\includegraphics[angle=0]{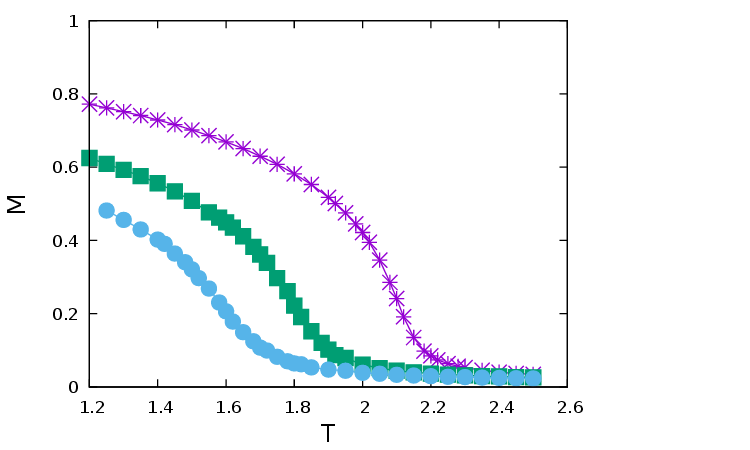}}

\caption{~Caption:~The magnetisation ($M$) is plotted against the Temperature ($T$) for different widths ($\omega$) of the  Gaussian distribution of the random anisotropy ($\gamma_{ij}$). Here $\omega=0.4$(Star), $\omega=0.9$(Box), $\omega=1.2$(Bullet).\\
Figure-16:~Alt-text:~The temperature dependence of the magnetisation for three
different widths of the Gaussian distribution of random anisotropy. Here also, the phase transition (where the magnetisation vanishes) takes place at lower
temperature for the higher value of the width of the Gaussian distribution of random anisotropy.
}
\label{M-T-different-gaussian-anis}
\end{center}
\end{figure}

\begin{figure}[h]
\begin{center}

%\resizebox{10cm}{!}{\includegraphics{Chi-T-different-gaussian-anis.pdf}}

\resizebox{17cm}{!}{\includegraphics[angle=0]{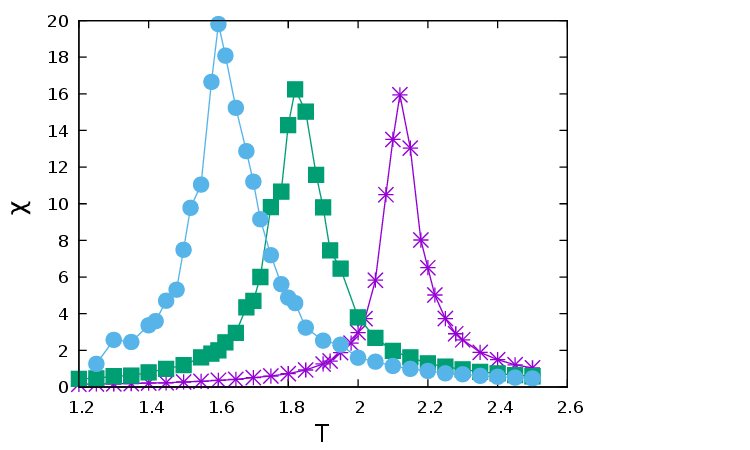}}

\caption{~Caption:~The susceptibility ($\chi$) is plotted against the Temperature ($T$) for different widths ($\omega$) of the Gaussian distribution of the random anisotropy ($\gamma_{ij}$). Here $\omega=0.4$(Star), $\omega=0.9$(Box), $\omega=1.2$(Bullet).\\
Figure-17:~Alt-text:~The temperature dependence of the susceptibility for three
different widths of the Gaussian distribution of random anisotropy. The susceptibility gets peaked (indicating the phase transition) at lower temperatures for
higher values of the width of the Gaussian distribution of random anisotropy.
}
\label{Chi-T-different-gaussian-anis}
\end{center}
\end{figure}

We have considered the Gaussian distribution (equation-\ref{normal}) of the
random anisotropy ($\gamma_{ij}$). Here also, we have observed the reduction of the zero temperature
magnetisation with the increase of the width of the Gaussian distribution of the anisotropy. We have considered three different values of the width 
($\omega = 0.4, 0.9, 1.2$).This has been shown in Fig-\ref{M-T-different-gaussian-anis}. The pseudocritical temperature was found from the position of the peak
of the susceptibility plotted against the temperature 
(Fig-\ref{Chi-T-different-gaussian-anis}). The pseudocritical temperature has been 
observed to decrease with the increase of the width of the distribution of the
anisotropy.

\noindent {\it {(iii) Bimodal distribution of random anisotropy}}

\begin{figure}[h]
\begin{center}

%\resizebox{10cm}{!}{\includegraphics{M-T-different-bimodal-anis.pdf}}

\resizebox{17cm}{!}{\includegraphics[angle=0]{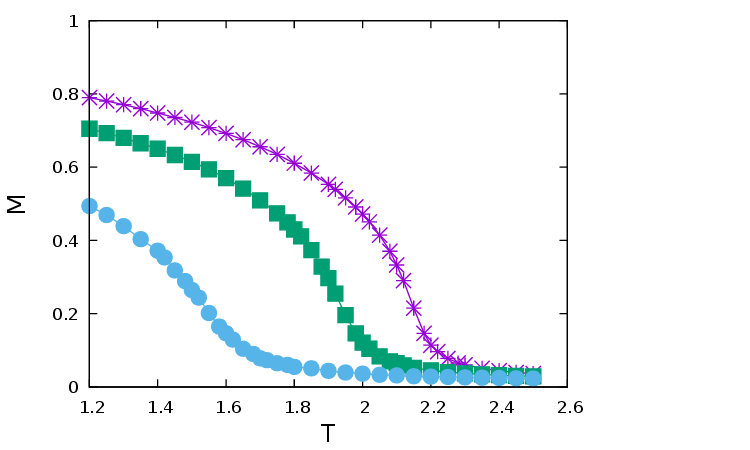}}

\caption{~Caption:~The magnetisation ($M$) is plotted against the Temperature ($T$) for different widths ($\omega$) of the  Bimodal distribution of the random anisotropy ($\gamma_{ij}$). Here $\omega=0.6$(Star), $\omega=1.2$(Box), $\omega=1.8$(Bullet).\\
Figure-18:~Alt-text:~Temperature dependence of the magnetisation for three different widths of the Bimodal distribution of the random anisotropy. The phase transition (where the magnetisation vanishes) takes place at lower temperatures for higher values of the width of the Bimodal
distribution of random anisotropy.}
\label{M-T-different-bimodal-anis}
\end{center}
\end{figure}

\begin{figure}[h]
\begin{center}

%\resizebox{10cm}{!}{\includegraphics{Chi-T-different-bimodal-anis.pdf}}

\resizebox{17cm}{!}{\includegraphics[angle=0]{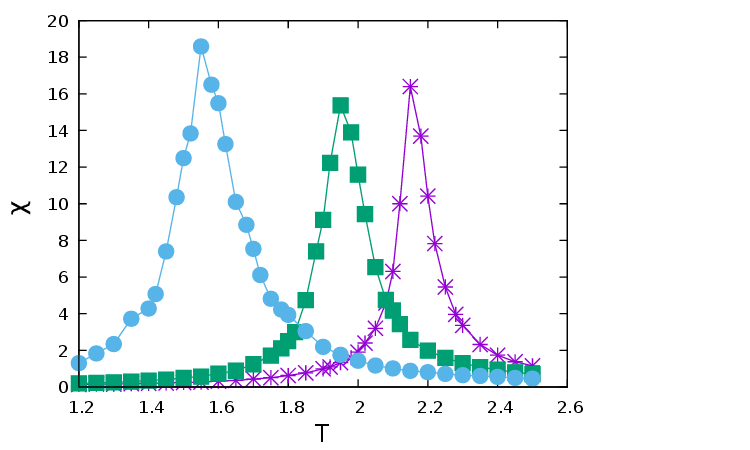}}

\caption{~Caption:~The susceptibility ($\chi$) is plotted against the temperature ($T$) for different widths ($\omega$) of the Bimodal distribution of the random anisotropy ($\gamma_{ij}$). Here $\omega=0.6$(Star), $\omega=1.2$(Box), $\omega=1.8$(Bullet).\\
Figure-19:~Alt-text:~The temperature dependence of the susceptibility for three
different widths of the Bimodal distribution of the random anisotropy. The susceptibility gets peaked (indicating the phase transition) at lower temperatures
for higher values of the width of the Bimodal distribution of random anisotropy.}
\label{Chi-T-different-bimodal-anis}
\end{center}
\end{figure}

The bimodal distribution (equation-\ref{bimodal}) of the random anisotropy ($\gamma_{ij}$) has been
studied.  We have noticed the reduction of the zero temperature
magnetisation with the increase of the width of the  distribution of the anisotropy. Three different values of the width
($\omega = 0.6, 1.2, 1.8$) have been considered here. 
The result has been represented pictorially in Fig-\ref{M-T-different-bimodal-anis}. The pseudocritical temperature has been estimated from the position of the peak
of the susceptibility plotted against the temperature 
(Fig-\ref{Chi-T-different-bimodal-anis}). The pseudocritical temperature was 
observed to decrease with the increase of the width of the distribution of the
anisotropy, as noticed in the previously discussed cases of distributed anisotropy.

\vskip 0.5cm

\noindent {\textbf {The phase boundary}}

\begin{figure}[h]
\begin{center}

%\resizebox{10cm}{!}{\includegraphics{phase-diagram-dist-anis.pdf}}

\resizebox{15cm}{!}{\includegraphics[angle=0]{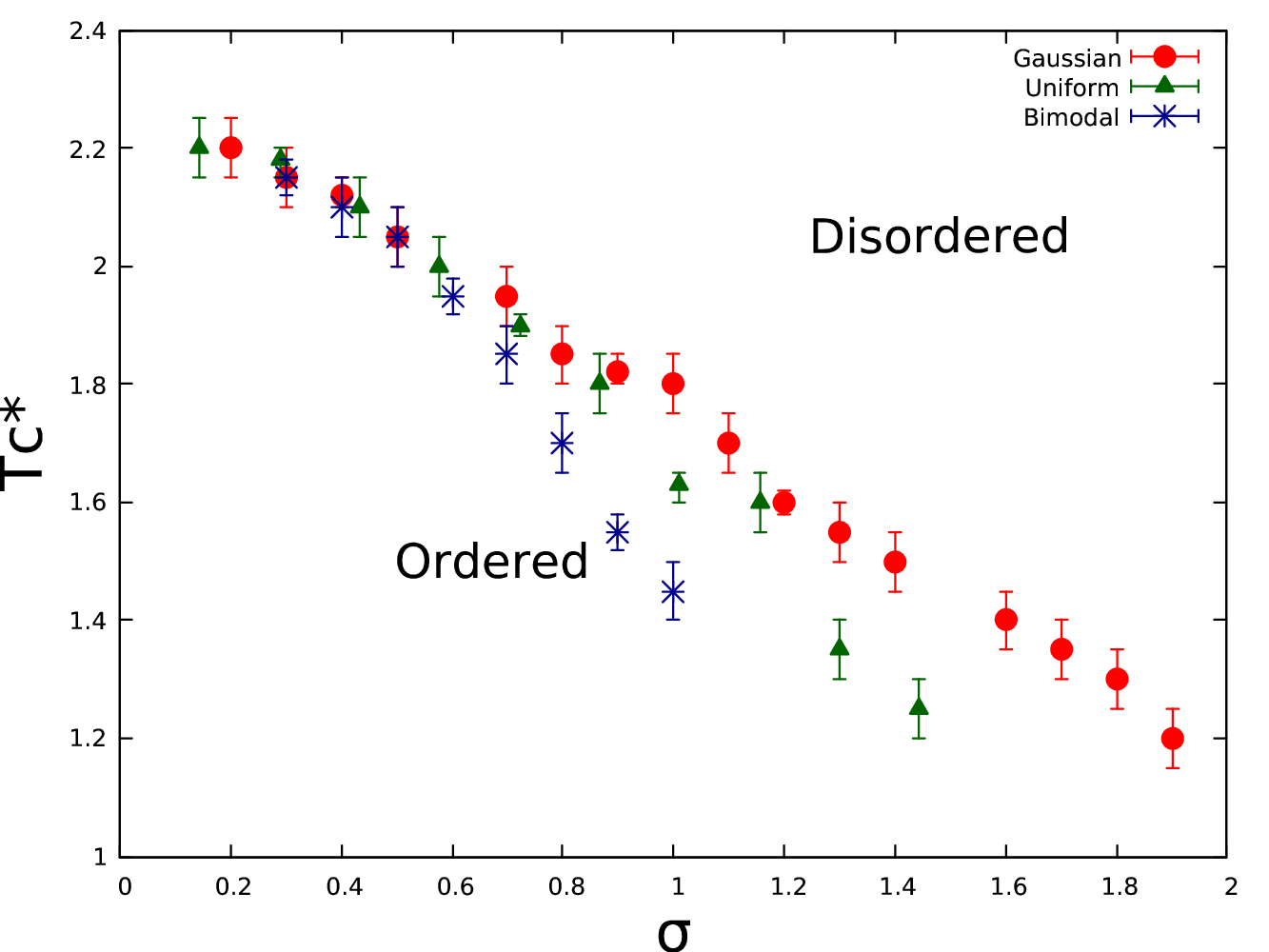}}

\caption{~Caption:~The pseudocritical temperature ($T_c^{*}$) is plotted against the standard deviations of the different distributions of the random  anisotropy ($\gamma_{ij}$):  Uniform(Green triangle), Gaussian(Red bullet), Bimodal(Blue star). The errorbars represent the maximum error involved in determining $T_c^{*}$.\\
Figure-20:~Alt-text:~The pseudocritical temperature versus the standard deviations of the different distribution of random anisotropy, are shown here. The pseudocritical temperature decreases as the standard deviation increases for all
kinds of distributions. The pseudocritical temperatures for different distributions of the random anisotropy are almost equal in the
limit of small values of the standard deviations. In the limit of vanishingly small value of the standard deviations (of all
types of the distribution), the pseudocritical temperature approaches the value of that for isotropic XY model in three dimensions. For larger value of the standard deviation, the pseudocritical temperatures are distinctly different for three
different kinds of distribution of random anisotropy.
}
\label{phase-diagram-dist-anis}
\end{center}
\end{figure}

The monotonic decrease of the pseudocritical temperature ($T_c^*$) with the 
increase of the width ($\omega$) of the distribution is noticed in the case
of distributed anisotropy. This functional dependence of $T_c^*(\omega)$, for
any particular kind of distribution, can be visualised as the phase boundary
of the phase transition. Below this boundary, one can have the ordered phase.
Since the standard deviation ($\sigma$) is linearly related to the width 
($\omega$) of the distribution of the anisotropy, this pseudocritical 
temperature is the function of the standard deviation ($\sigma$) also. The 
phase boundary is shown in Fig-\ref{phase-diagram-dist-anis} for three different
kinds of distribution of the anisotropy. It may be noted here  that for vanishingly small standard deviation ($\sigma \to 0$) of the distributed anisotropy, the pseudocritical temperatures become insensitive to the nature of the distribution and
approaches the critical value \cite{univ,hasenbusch}
of that of the {\it isotropic} XY ferromagnet in 3D.

\vskip 1cm

\vskip 0.5cm

\begin{center}{\bf Table-I}\\

\vskip 0.2cm

\begin{tabular}{ |p{3cm}|p{3cm}|p{3cm}|   }
\hline
\multicolumn{3}{|c|}{Scaling exponents} \\
\hline
Type &${\beta\over\nu}$ &${\gamma\over\nu}$ \\
\hline
Constant-anisotropy & $0.271\pm 0.010$ &$1.751\pm 0.162$ \\
\hline
Uniformly-distributed-random-anisotropy & $0.510\pm 0.015 $ & $1.863\pm0.033$  \\
\hline

\end{tabular}

\end{center} 

\noindent {Table-I. The values of the critical exponents for constant anisotropy and for uniformly distributed random anisotropy}
\vskip 0.8cm

\noindent {\bf V. Concluding remarks:}

In this paper, we have systematically investigated the critical behaviours of a three-dimensional anisotropic  XY ferromagnet by extensive Monte Carlo simulation using Metropolis algorithm. The change in the pseudocritical temperature due to the presence of anisotropy is the main objective of our study. The critical temperature of the isotropic three-dimensional XY ferromagnet was already reported \cite{hasenbusch} for Monte Carlo simulation. We have studied the case of constant anisotropy here. Our numerical results show that the pseudocritical temperature depends on the strength of the constant anisotropy. As the strength of the anisotropy increases, the pseudocritical temperature also increases linearly. This behaviour can be qualitatively explained as follows: for a nonzero strength of the constant anisotropy, the system has a relative bias in the form of ordering. This enhanced biassing of the ordering would require more thermal energy to break into the disordered phase. As a result, the pseudocritical temperature increases. However, a completely different scenario has been observed in the case where the anisotropy is random but statistically distributed. We have considered specific statistical distributions of the random anisotropy in such a manner that in all cases, the mean strength of the anisotropy is zero. We have considered three different kinds of statistical distribution of the random anisotropy, namely,  uniform distribution, Gaussian distribution, and bimodal distribution. In all the cases, the pseudocritical temperatures are observed to decrease as the widths (and the standard deviations) of the distributions increased. In the case of distributed anisotropy, the different spin-spin interactions experience different values of the strength of the anisotropy. As a result, the system cannot be biased in the form of ordering. Hence, the distributed anisotropy plays the role of a disordering field and causes the reduction of pseudocritical temperatures.
We have drawn the phase boundaries (for three different kinds of distribution of the anisotropy) as shown in Fig-\ref{phase-diagram-dist-anis}. 

The finite size analysis with fourth order Binder cumulant has been done for some cases to determine the critical temperature in the thermodynamic limit. Following the scaling relations $M \sim L^{-{\beta\over\nu}}$ and $\chi \sim L^{{\gamma\over\nu}}$, the exponents are estimated and provided in Table-I. The anisotropic XY ferromagnet
belongs to the different universality class of isotropic XY ferromagnet.
Moreover, the randomly anisotropic XY ferromagnet belongs to a different universality class than that of constant anisotropic XY ferromagnet. 

There are some important issues which can be studied further as suggested by the reviewers of this manuscript. The dependence
of the critical exponents on the anisotropy may lead to the multifractal behaviour. This should be checked rigorously to conclude regarding the multifractality (if any). The temperature dependence of the
specific heat may be studied in detail and corresponding critical exponent $\alpha/{\nu}$ (if any scaling exists $C_v \sim L^{-\alpha/{\nu}}$) may be estimated. The study of the correlation function is also important. All these studies require huge computational efforts
of several months. These may be studied as a different follow up project and the results may be published elsewhere. In our overall calculations the total CPU times required is approximately 673 hours
For further calculations of correlation functions and specific heat (precisely near the transition point both from the fluctuation of energy as well as the derivative of energy.), the estimated and approximate CPU time would be 267 hours in the HP intel XEON processor. 

\vskip 0.5cm

\noindent {\bf Acknowledgements:} We thank Diptiman Sen, IISc, Bangalore, for discussion. OM acknowledges MANF,UGC, Govt. of India for financial support. We thank Ishita Tikader for a careful reading 
of the manuscript. We thank the anonymous reviewers for some important suggestions.

\vskip 1cm

\noindent {\bf Data availability statement:} Data will be available on request to
Olivia Mallick.

\vskip 0.5cm

\noindent {\bf Conflict of interest statement:} We declare that this manuscript is
free from any conflict of interest. The authors have no financial or proprietary interests in any material discussed in this article.

\vskip 0.5cm

\noindent {\bf Funding statement:} No funding was received particularly to support this work.

\vskip 0.5cm
\noindent {\bf Authors' contributions:}  Olivia Mallick-developed the code, collected the data, prepared the figures, wrote the manuscript. Muktish Acharyya-conceptualized the problem, developed the code, analysed the results, wrote the manuscript.
\newpage

\bibliographystyle{unsrt}
\bibliography{olivia-PT}

\newpage

\newpage

%%%%%%%%%%%%%%%%%%%%%%%%%%%%%%%%%%%%%%%%%%%%%%%%%%%%%%%%%%%%%%%%%%%%%%%%%%%%%%%%
%%%%%%%%%%%%%%%%%%%%%%%%%%%%%%%%%%%%%%%%%%%%%%%%%%%%%%%%%%%%%%%%%%%%%%%%%%

\newpage
%%%%%%%%%%%%%%%%%%%%%%%%%%%%%%%%%%%%%%%%%%%%%%%%%%%%%%%%%%%%%%%%%%%%%%%%%%%%%%%%
%%%%%%%%%%%%%%%%%%%%%%%%%%%%%%%%%%%%%%%%%%%%%%%%%%%%%%%%%%%%%%%%%%%%%%%%%%

\newpage
%%%%%%%%%%%%%%%%%%%%%%%%%%%%%%%%%%%%%%%%%%%%%%%%%%%%%%%%%%%%%%%%%%%%%%%%%%%%%%%%
%%%%%%%%%%%%%%%%%%%%%%%%%%%%%%%%%%%%%%%%%%%%%%%%%%%%%%%%%%%%%%%%%%%%%%%%%%

\newpage
%%%%%%%%%%%%%%%%%%%%%%%%%%%%%%%%%%%%%%%%%%%%%%%%%%%%%%%%%%%%%%%%%%%%%%%%%%%%%%%%
%%%%%%%%%%%%%%%%%%%%%%%%%%%%%%%%%%%%%%%%%%%%%%%%%%%%%%%%%%%%%%%%%%%%%%%%%%

\newpage
%%%%%%%%%%%%%%%%%%%%%%%%%%%%%%%%%%%%%%%%%%%%%%%%%%%%%%%%%%%%%%%%%%%%%%%%%%%%%%%%
%%%%%%%%%%%%%%%%%%%%%%%%%%%%%%%%%%%%%%%%%%%%%%%%%%%%%%%%%%%%%%%%%%%%%%%%%%

\newpage
%%%%%%%%%%%%%%%%%%%%%%%%%%%%%%%%%%%%%%%%%%%%%%%%%%%%%%%%%%%%%%%%%%%%%%%%%%%%%%%%
%%%%%%%%%%%%%%%%%%%%%%%%%%%%%%%%%%%%%%%%%%%%%%%%%%%%%%%%%%%%%%%%%%%%%%%%%%of the pseudocritical temperature on 

\newpage
%%%%%%%%%%%%%%%%%%%%%%%%%%%%%%%%%%%%%%%%%%%%%%%%%%%%%%%%%%%%%%%%%%%%%%%%%%%%%%%%
%%%%%%%%%%%%%%%%%%%%%%%%%%%%%%%%%%%%%%%%%%%%%%%%%%%%%%%%%%%%%%%%%%%%%%%%%%

\newpage
%%%%%%%%%%%%%%%%%%%%%%%%%%%%%%%%%%%%%%%%%%%%%%%%%%%%%%%%%%%%%%%%%%%%%%%%%%%%%%%%
%%%%%%%%%%%%%%%%%%%%%%%%%%%%%%%%%%%%%%%%%%%%%%%%%%%%%%%%%%%%%%%%%%%%%%%%%%
\newpage
%%%%%%%%%%%%%%%%%%%%%%%%%%%%%%%%%%%%%%%%%%%%%%%%%%%%%%%%%%%%%%%%%%%%%%%%%%%%%%%%
%%%%%%%%%%%%%%%%%%%%%%%%%%%%%%%%%%%%%%%%%%%%%%%%%%%%%%%%%%%%%%%%%%%%%%%%%%
\newpage
%%%%%%%%%%%%%%%%%%%%%%%%%%%%%%%%%%%%%%%%%%%%%%%%%%%%%%%%%%%%%%%%%%%%%%%%%%%%%%%%
%%%%%%%%%%%%%%%%%%%%%%%%%%%%%%%%%%%%%%%%%%%%%%%%%%%%%%%%%%%%%%%%%%%%%%%%%%

\newpage
%%%%%%%%%%%%%%%%%%%%%%%%%%%%%%%%%%%%%%%%%%%%%%%%%%%%%%%%%%%%%%%%%%%%%%%%%%%%%%%%
%%%%%%%%%%%%%%%%%%%%%%%%%%%%%%%%%%%%%%%%%%%%%%%%%%%%%%%%%%%%%%%%%%%%%%%%%%
\newpage
%%%%%%%%%%%%%%%%%%%%%%%%%%%%%%%%%%%%%%%%%%%%%%%%%%%%%%%%%%%%%%%%%%%%%%%%%%%%%%%%
%%%%%%%%%%%%%%%%%%%%%%%%%%%%%%%%%%%%%%%%%%%%%%%%%%%%%%%%%%%%%%%%%%%%%%%%%%

\newpage
%%%%%%%%%%%%%%%%%%%%%%%%%%%%%%%%%%%%%%%%%%%%%%%%%%%%%%%%%%%%%%%%%%%%%%%%%%%%%%%%
%%%%%%%%%%%%%%%%%%%%%%%%%%%%%%%%%%%%%%%%%%%%%%%%%%%%%%%%%%%%%%%%%%%%%%%%%%

\newpage
%%%%%%%%%%%%%%%%%%%%%%%%%%%%%%%%%%%%%%%%%%%%%%%%%%%%%%%%%%%%%%%%%%%%%%%%%%%%%%%%
%%%%%%%%%%%%%%%%%%%%%%%%%%%%%%%%%%%%%%%%%%%%%%%%%%%%%%%%%%%%%%%%%%%%%%%%%%

\newpage
%%%%%%%%%%%%%%%%%%%%%%%%%%%%%%%%%%%%%%%%%%%%%%%%%%%%%%%%%%%%%%%%%%%%%%%%%%%%%%%%
%%%%%%%%%%%%%%%%%%%%%%%%%%%%%%%%%%%%%%%%%%%%%%%%%%%%%%%%%%%%%%%%%%%%%%%%%%

\newpage
%%%%%%%%%%%%%%%%%%%%%%%%%%%%%%%%%%%%%%%%%%%%%%%%%%%%%%%%%%%%%%%%%%%%%%%%%%%%%%%%
%%%%%%%%%%%%%%%%%%%%%%%%%%%%%%%%%%%%%%%%%%%%%%%%%%%%%%%%%%%%%%%%%%%%%%%%%%
\newpage
%%%%%%%%%%%%%%%%%%%%%%%%%%%%%%%%%%%%%%%%%%%%%%%%%%%%%%%%%%%%%%%%%%%%%%%%%%%%%%%%
%%%%%%%%%%%%%%%%%%%%%%%%%%%%%%%%%%%%%%%%%%%%%%%%%%%%%%%%%%%%%%%%%%%%%%%%%%
\newpage
%%%%%%%%%%%%%%%%%%%%%%%%%%%%%%%%%%%%%%%%%%%%%%%%%%%%%%%%%%%%%%%%%%%%%%%%%%%%%%%%
%%%%%%%%%%%%%%%%%%%%%%%%%%%%%%%%%%%%%%%%%%%%%%%%%%%%%%%%%%%%%%%%%%%%%%%%%%
\newpage
%%%%%%%%%%%%%%%%%%%%%%%%%%%%%%%%%%%%%%%%%%%%%%%%%%%%%%%%%%%%%%%%%%%%%%%%%%%%%%%%
%%%%%%%%%%%%%%%%%%%%%%%%%%%%%%%%%%%%%%%%%%%%%%%%%%%%%%%%%%%%%%%%%%%%%%%%%%

\newpage
%%%%%%%%%%%%%%%%%%%%%%%%%%%%%%%%%%%%%%%%%%%%%%%%%%%%%%%%%%%%%%%%%%%%%%%%%%%%%%%%
%%%%%%%%%%%%%%%%%%%%%%%%%%%%%%%%%%%%%%%%%%%%%%%%%%%%%%%%%%%%%%%%%%%%%%%%%%

\end{document}